\documentclass[conference]{IEEEtran}
\usepackage{cite}
\usepackage{amsmath,amssymb,amsfonts}
\usepackage{algorithmic}
\usepackage{graphicx}
\usepackage{textcomp}
\usepackage{xcolor}
\usepackage{fancyhdr}
\usepackage[hyphens]{url}
\usepackage{cite}
\usepackage[inline]{enumitem}
\usepackage{booktabs}
\usepackage{multirow}
\usepackage{dsfont}

\usepackage{amsmath}
\usepackage{amsfonts,bm}
\usepackage{outlines}
\usepackage{textcomp}
\usepackage{graphicx}
\usepackage{multicol}

\usepackage{subcaption}
\def\BibTeX{{\rm B\kern-.05em{\sc i\kern-.025em b}\kern-.08em
    T\kern-.1667em\lower.7ex\hbox{E}\kern-.125emX}}

\usepackage{comment}

\fancypagestyle{firstpage}{
  \fancyhf{}

  \fancyhead[C]{} 
  \fancyfoot[C]{\thepage}
}  

\title{LAORAM: A Look Ahead ORAM Architecture for Training Large Embedding Tables} 

 \author{\IEEEauthorblockN{Rachit Rajat\IEEEauthorrefmark{1}}
 \IEEEauthorblockA{University of Southern California\\
 \texttt{rrajat@usc.edu}
 \thanks{\IEEEauthorrefmark{1}Equal contribution}}
 \and
 \IEEEauthorblockN{Yongqin Wang\IEEEauthorrefmark{1}}
 \IEEEauthorblockA{University of Southern California\\
 \texttt{yongqin@usc.edu }}
 \and
 \IEEEauthorblockN{Murali Annavaram}
 \IEEEauthorblockA{University of Southern California\\ 
\texttt{annavara@usc.edu }}}

\begin{document}
\maketitle
\thispagestyle{firstpage}
\pagestyle{plain}
\def\thefootnote{*}\footnotetext{Equal Contribution}

\begin{abstract}
\label{sec:abstract}
Data confidentiality/privacy is becoming a significant concern, especially in the cloud computing era. Memory access patterns have been demonstrated to leak critical information such as security keys and a program's spatial and temporal information. This information leak poses an even more significant privacy challenge in machine learning models with embedding tables. Embedding tables are routinely used to learn categorical features from training data. Even knowing the locations of the embedding table entries accessed, not the data within the embedding table, will compromise categorical input data to the model. Embedding entries are privacy sensitive since they disclose valuable properties about the user. Oblivious RAM (ORAM), and its enhanced variants such as PathORAM have emerged as viable solutions to hide leakage from memory access streams. PathORAM fetches an entire path of memory blocks even if a single block is needed. Once the block is fetched a new path is randomly assigned thereby leading to substantial bandwidth and performance overheads. 

In this work, we present LAORAM, an ORAM framework explicitly designed to protect user privacy during embedding table training. LAORAM exploits the unique property of training, namely the training samples that are going to be used in the future are known beforehand. LAORAM preprocesses the training samples (securely without revealing the entry values) to identify the memory blocks which are accessed together in the near future. The system tries to assign these blocks to as few paths as possible within the PathORAM infrastructure. 

LAORAM does this operation by combining multiple blocks accessed together as superblocks. Thus, future accesses to a collection of blocks can be satisfied from a few paths, effectively reducing the number of reads and writes required by the framework. To further increase performance, LAORAM uses a fat-tree structure for PathORAM, i.e. a tree with variable bucket size, effectively reducing the number of background evictions required, which improves the stash usage. We have evaluated LAORAM using both a recommendation model (DLRM) and a NLP model (XLM-R) embedding table configurations. LAORAM performs 5 times faster than PathORAM on a recommendation dataset (Kaggle) and 5.4x faster on a NLP dataset (XNLI), while guaranteeing the same security guarantees as the original PathORAM.

\end{abstract}

\section{Introduction}
\label{sec:introduction}
Privacy of user data, particularly in the context of using cloud resources, is becoming critical. In the context of machine learning, data owners want to keep their data confidential when using cloud vendors like Amazon AWS or Microsoft Azure to train or infer from their models. Privacy preserving machine learning has gained prominence in recent years. Researchers have explored different approaches to privacy protection in machine learning. Their goal has been to primarily protect the data and/or model parameters from being visible to an adversary. Prior approaches to protect data privacy include Homomorphic Encryption~\cite{gentry2009fully,liu2017oblivious,gilad2016cryptonets,juvekar2018gazelle}, Secure Multi-Party Computing (MPC)~\cite{shokri2015privacy, mohassel2017secureml, wagh2019securenn, mohassel2018aby3, wang2022charact}, Differential Privacy~\cite{abadi2016deep, erlingsson2014rappor} and using Trusted Execution Enviroments~\cite{TensorSCONE, tramer2019slalom, darkNight, origami}.

While these prior schemes provide privacy of data or model, there is a class of machine learning workloads which use embedding tables, where privacy requires protecting embedding table entry addresses accessed by the application, even when the embedding table content is protected. Embedding tables will use a unique row in the table (an embedding vector) to represent a unique categorical data (one-to-one mapping, see figure~\ref{fig:attack}). Hence, even knowing the embedding table entry that is being accessed, which is a memory address, can compromise privacy.  Embedding tables are increasingly at the core of multiple high valued machine learning models, such as recommendation models~\cite{dlrm, dp-in-ali, alirs} and transformer-based natural language processing (NLP) models~\cite{bert, roberta, xlm, xlmr}. Many of the inputs to embedding tables are categorical information associated with a user. For instance in recommendation systems, embedding table entries capture a user's prior behavior, such as advertisement clicking choices and browsed news feeds. In the case of NLP models, each embedding entry may be associated with a learned representation of a word. Thus the addresses associated with accessing categorical data in an embedding table reveal a lot about the user. Thus, ML models using embedding tables require protection against memory address leakage.


\subsection{What do Embedding Tables Reveal?}
One natural question that arises is what do embedding table entries reveal? Embedding tables can convert categorical data to a more expressive data format. Categorical data can be difficult for machine learning systems to handle because semantically similar items in most cases are not close in the form of sparse vectors. Embedding tables will learn to map semantically similar inputs closer in the embedding space during training. 


A training sample for embedding tables consists of $t$ embedding table entries  that must be accessed. The embedding table lookup for a training sample is defined as $R = A * W$, where $W$ is the embedding table and $A= [e_1, e_2, ..., e_t]$. Each $e_i$ is a one-hot vector representing categorical inputs to embedding tables. As the size of the embedding table is enormous (terabytes)~\cite{aibox-one-node-training}, practically, embedding table accesses are implemented as fetching embedding entries corresponding to each $e_i$. 
The entry number of an accessed embedding table entry  $e_i$ is a valuable training sample that holds sensitive information. 

\begin{figure}
\centering
  \includegraphics[width=8cm]{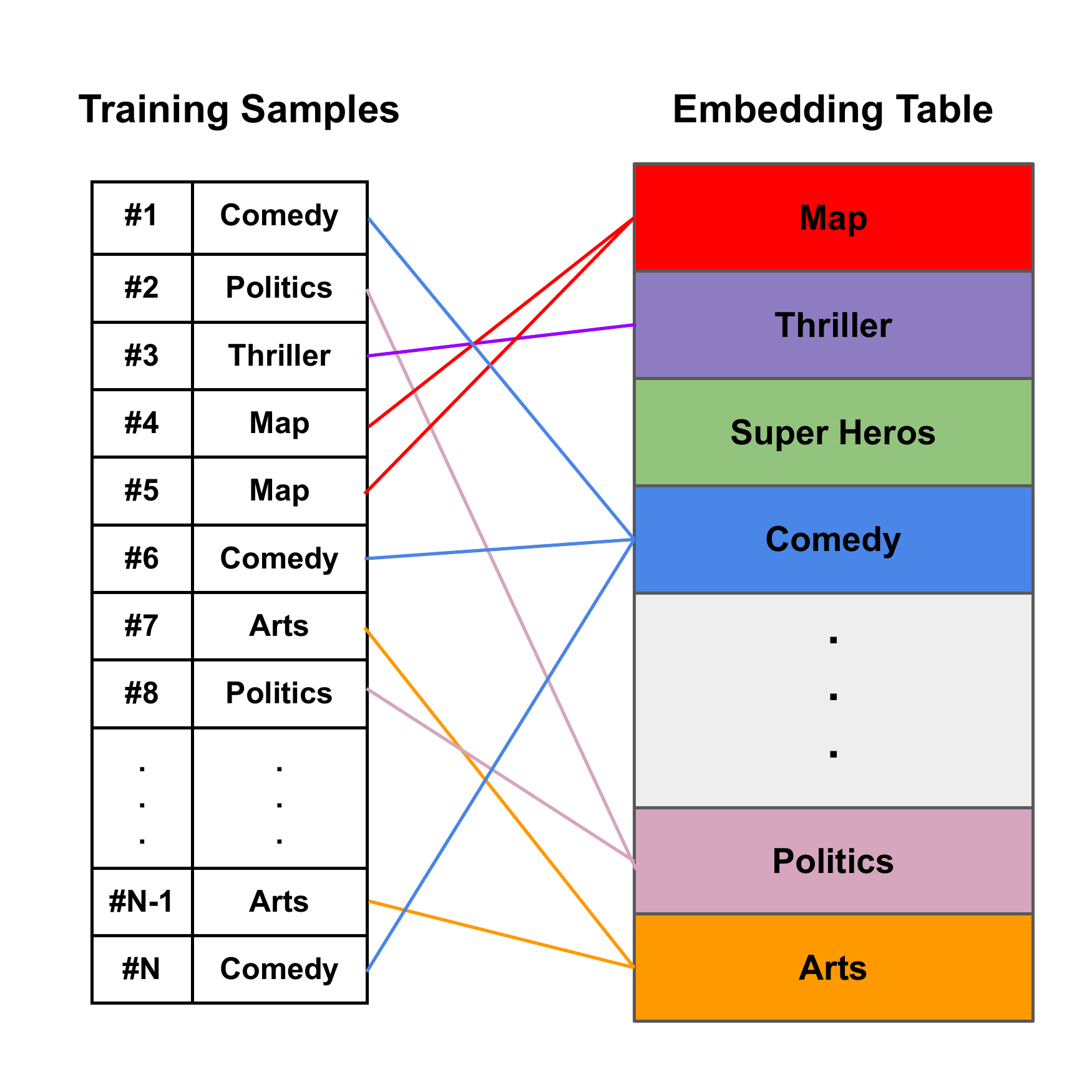}\\
  \caption{Left: $N$ training samples representing most watched video types;Right: a embedding table; Each training sample needs to fetch one entry of the embedding table.}
  \label{fig:attack}
\end{figure}

Consider an example training sequence shown in figure \ref{fig:attack}. The figure shows an embedding table that captures several different types of videos watched by users. In this example, the first training sample shows that a user watches \textit{comedy} movies the most. The second sample shows that a user watches  \textit{political} movies the most. The next user watches \textit{thrillers} and so on. Let's assume users who watches comedies the most will likely to watch super hero movies. After embedding table training, embedding table entries representing those two categories are likely to be similar.

What is critical to note in this example is that each of the movie categories correspond to a specific embedding table entry. Hence, knowing the embedding table entry (which is a memory address) can compromise the user's movie preference. While a user's movie preferences may look innocuous, this example can easily extend to more challenging situations where a user's political affiliation or web surfing behavior can be extracted from knowing the embedding table entries.  
A curious OS can utilize a combination of page faults with cache side channel snooping to obtain embedding table access patterns at a cache line granularity. In particular, a curious OS can set all present bits of embedding table memory pages to $0$. Consequently, every time the embedding table is accessed, a page fault is triggered. The OS can know the page number being accessed from the information provided to the page fault handler. After handling the page fault, the OS can use \textbf{flush+reload} attacks to get the accessed address at a cache line granularity. Thus if each cache line corresponds to one embedding entry then the user's private information is compromised. 

\subsection{Memory Address Leakage Mitigation}
The most popular mitigation technique for hiding memory access patterns is Oblivious RAM (ORAM)\cite{original-oram}. Conceptually, ORAMs access a large number of data blocks for every memory access request to obfuscate the actual data address. Furthermore, ORAM reshuffles data placement so that future accesses to the same address will generate a different random set of addresses on the bus. ORAM guarantees that the adversary cannot distinguish between two memory accesses. Recent enhancements to  ORAM design include Path ORAM~\cite{path-oram}, where the data blocks are stored in the form of a tree, and each data block is assigned a path (i.e. a leaf node). On a data block access, not only the requested data block is read but also all the blocks along the path from the root node to the leaf node are read (more in-depth detail in the background section). Unfortunately, the security guarantees of nearly all the ORAMs \cite{path-oram, ring-oram, original-oram, streamline-ring-oram} come at the cost of significant overhead. Theorem 1.2.2 in \cite{original-oram} indicates that when accessing $t$ data blocks of a RAM which contains $N$ data blocks, additional $max\{N, (t-1)\cdot log(N)\} - t$ data blocks need fetching in ORAM. Even the optimized PathORAM\cite{path-oram} needs to access $t\cdot[log(N) - 1]$ data blocks for each block accessed. These additional accesses cause memory access delays and bandwidth contention. Besides memory access delays, ORAM metadata management such as stash management, position map lookups, and data eviction logic add additional overheads for each ORAM access.

Proposals have been made to reduce the overheads of PathORAM, such as using either static or dynamic superblocks as described in PrORAM~\cite{proram}. Superblocks reduce the number of paths that need to be read by coalescing multiple memory accesses to be accessed by the application into a single path. Instead of reading $n$ paths for $n$ data accesses, a \textit{perfectly} formed superblock of size $S$ needs to read $n/S$ paths from ORAM. Prior techniques strive to create such superblocks \cite{superblock-pathOram} by merging adjacent data blocks into bigger blocks. For example, PrORAM~\cite{proram} groups \textbf{adjacent} blocks that were previously accessed together into a few superblocks with the hope that these blocks are likely to be accessed together in the future. 

\begin{figure}[h]
\centering
  \includegraphics[width=8cm]{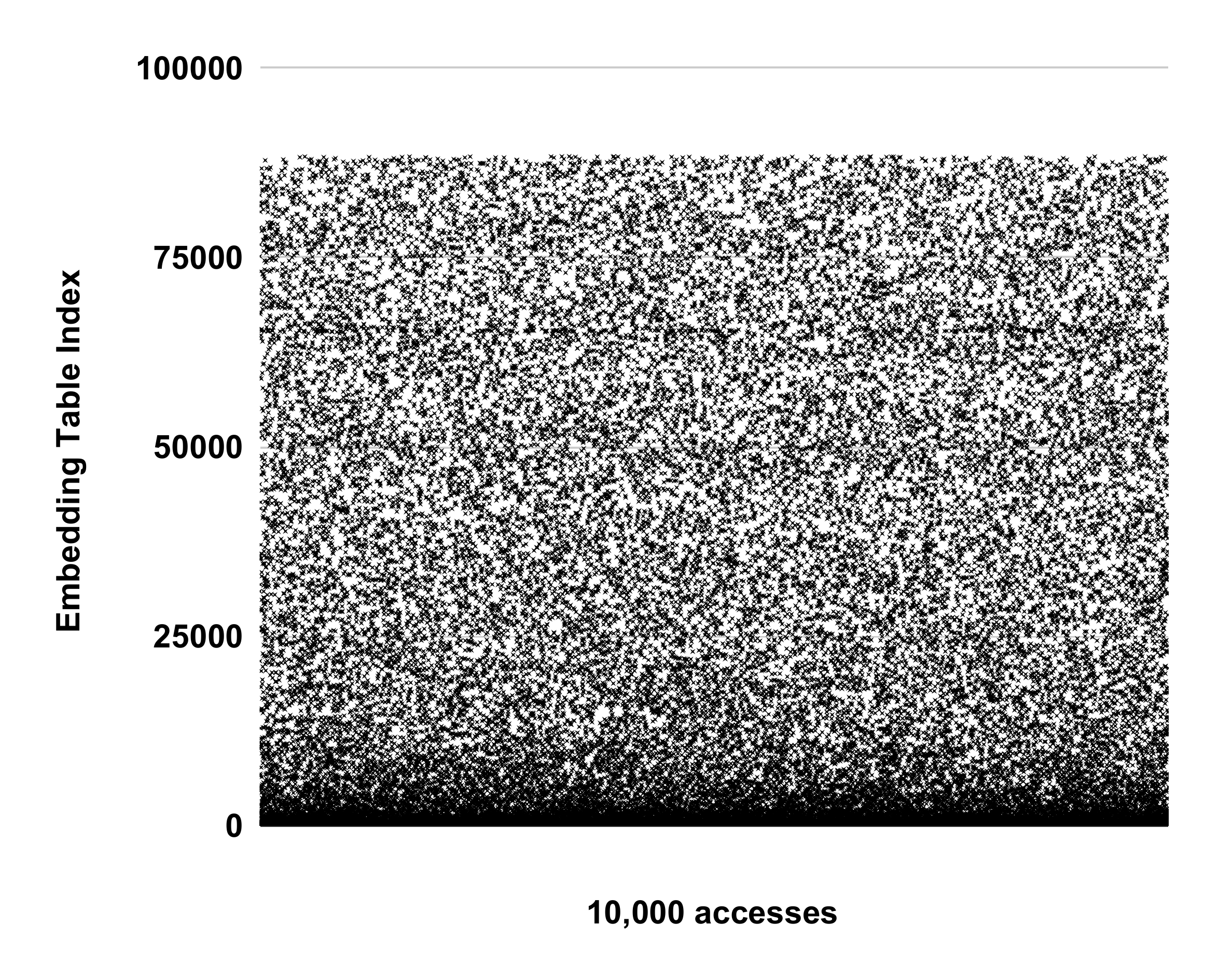}\\
  \caption{10000 accesses to embedding tables for Kaggle Dataset.}
  \label{fig:random-access}
\end{figure}

Looking at the past accesses to  form  the superblocks is very challenging for embedding table training systems. Figure~\ref{fig:random-access} plots the embedding table entries that are accessed at each training sample for the first 10,000 training samples for the Kaggle dataset, which is provided as a representative dataset for the DLRM recommendation model by Meta Inc~\cite{dlrm}. One can clearly see that most accesses are random, and only a  narrow black band at the bottom of the figure illustrates that a few indices are accessed repeatedly. In the absence of good predictability, PrORAM performs similarly to the  PathORAM. 

Unfortunately, the embedding table accesses across different training samples exhibit very little predictability. However, machine learning training exhibits a unique property. The training samples used for the several subsequent batches are known beforehand. Although predicting the future based on the past may be difficult, peeking into the future is feasible with training data. Thus future data access pattern is known beforehand. Peeking into the future provides oracular capabilities that allow LAORAM to form superblocks more aggressively. Unlike PrORAM, Look Ahead ORAM (LAORAM) takes advantage of this future knowledge to improve the performance of address hiding schemes. 

\subsection{Our Contributions}
\textbf{1. An aggressive dynamic superblock technique}: In this paper, we propose LAORAM, an aggressive dynamic superblock formation technique that takes advantage of the future access patterns to  reduce the number of reads/writes and access runtime. In LAORAM, the data blocks accessed together in upcoming training batches are combined to form superblocks. The future access patterns are discerned by using a low overhead preprocessing of the training samples. This look ahead dynamic superblock scheme reduces the number of reads/writes required, speeds up the memory access latency, and reduces the bandwidth requirement.

\textbf{2. Fat-tree structures for reduced stash demand}: Besides the usage of superblocks, we also propose to use fat-tree ORAM organization to mitigate the increased stash usage due to superblocks. Stash is a trusted memory chunk within the ORAM client, and accesses to the stash memory are assumed to be invisible in all ORAM designs. When using a superblock-based ORAM design, the ORAM must have sufficient empty blocks along the accessed path (more in the section~\ref{sec:background}) for the superblock. The superblocks may need to be temporarily stashed on the client in the absence of such empty blocks. Our analysis showed that if the superblock size is more than two blocks, the stash size grows rapidly. The reason for large stash growth is that if the superblock size is large, every time a data block gets a new address, there is little space in the write-back path to fit the accessed block. Hence, the newly formed superblock may have to wait in the stash for the availability of free space in the write-back path, which causes the stash to grow dramatically. Prior studies~\cite{proram} has also observed this fact.

A simple solution to stash growth is to increase the bucket size to enable writing large superblocks. A bigger bucket size, however, increases the memory space requirement. To solve this conundrum, LAORAM proposes a unique change to PathORAM. Instead of using a fixed bucket size for the entire tree, it uses variable bucket sizes similar to what a fat-tree structure\cite{fat} would entail. In this fat-tree structure, the node at the root has the largest bucket size. Nodes' bucket sizes will decrease closer to the leaves. This technique can significantly reduce stash usage when the superblock size is large.

Finally, we have evaluated LAORAM using both a recommendation model (DLRM) and a NLP model (XLM-R) embedding table configuration. LAORAM performs 5 times faster on the recommendation dataset (Kaggle) and 5.4x on a NLP dataset (XNLI).

\subsection{Scope of the LAORAM}
Our work is applicable to machine learning training systems that use embedding tables, such as recommendation systems, NLP models, and even image and video classification systems that are driven by Transformers~\cite{VideoBERT}. These models are the key enablers to growth in many industries. For instance, nearly 80\% of the inference cycles in current data centers are spent in recommendation models~\cite{chui2018notes}. Embedding table based NLP models power nearly all the voice enabled smart speakers in the world. And, nearly 70\% of the smart speaker users use voice assistants to access network resources.~\cite{amazonEcho70}. Similarly, current video streaming services are no longer video-on-demand systems, instead they are videos automatically recommended to the users~\cite{netflixResearch}, which in turns is driven by embedding tables. Thus protecting privacy of embedding table access patterns is a key requirement for sustained future growth of the industry. 






\section{Background}
\label{sec:background}
\subsection{CPU-centric Large Embedding Tables}
Large embedding tables can be trained on a variety of computing platforms. For instance, one way to train the  DLRM (a huge recommendation model from Meta)~\cite{dlrm} is to distribute the huge embedding tables across many GPUs~\cite{mudigere2021high}. Since each GPU has limited memory, the model must be distributed across hundreds of GPUs. Recent work \cite{aibox-one-node-training, cdlrm} showed that in these distributed training systems, GPUs are vastly underutilized, and using many GPUs primarily to store the model in their memory is expensive. Hence, there is a trend towards keeping large embedding tables on a CPU DRAM where the CPU fetches the embedding tables from DRAM and then feeds those entries to the GPU for training. Accessing the embedding tables in the CPU DRAM is a potential source of address leakage. Thus, while providing cost-efficient solutions, CPU-centric embedding tables also bring the address leakage problem to the forefront. 

\subsection{Oblivious RAM}
ORAM \cite{original-oram} uses the notion of a client and server. The data is stored on the server memory, and the client requests data from the server. The server accesses the data at the address sent by the client and returns it to the client. In this framework, the data stored in the server could be encrypted, and hence the only information leakage that occurs is the memory address patterns. The goal of the ORAM  framework is to protect against adversaries that can launch attacks on the server to observe memory access patterns. When ORAM is used, an adversary cannot figure out which address is being accessed even if they can observe the address stream. In other words, ORAM will not leak information about which data block is accessed, whether the same data block is being reaccessed and whether the access is a read or a write. Moreover, given two data request sequences of the same length, ORAM guarantees that their access patterns are computationally indistinguishable. 

\subsection{PathORAM}
PathORAM is an implementation of the ORAM when a client is assumed to have a small amount of private storage, called the stash \cite{path-oram}. PathORAM has two major components: the binary tree storage, which is on the server-side, and the ORAM stash controller, which is on the client-side. The stash controller consists of small trusted storage called stash and a position map data structure.
\textbf{Binary Tree Storage:} 
The data blocks in PathORAM are stored in the form of a binary tree. Each node can hold up to $Z$ data blocks, and we call $Z$ the bucket size of the node. Of these Z data blocks in each tree node, a subset of blocks (could be all Z blocks) contains real user data, while the rest of the blocks (dummy blocks) contains dummy data. The root node is considered level 0, while the leaf is considered level L. Each leaf has a path number associated with it.

\textbf{Client Side:}
The stash controller on the client consists of a stash and a position map. The stash is a small storage that holds data if the data blocks cannot be written back into the binary tree storage due to limited empty block availability within a given path. The position map is a lookup table that maps each real data block to one of the paths in the binary tree. A data block will be either in the binary tree storage or stash. The client's accesses to the stashed data are invisible to any adversary. 

\begin{figure}[h]
\centering
  \includegraphics[width=8cm]{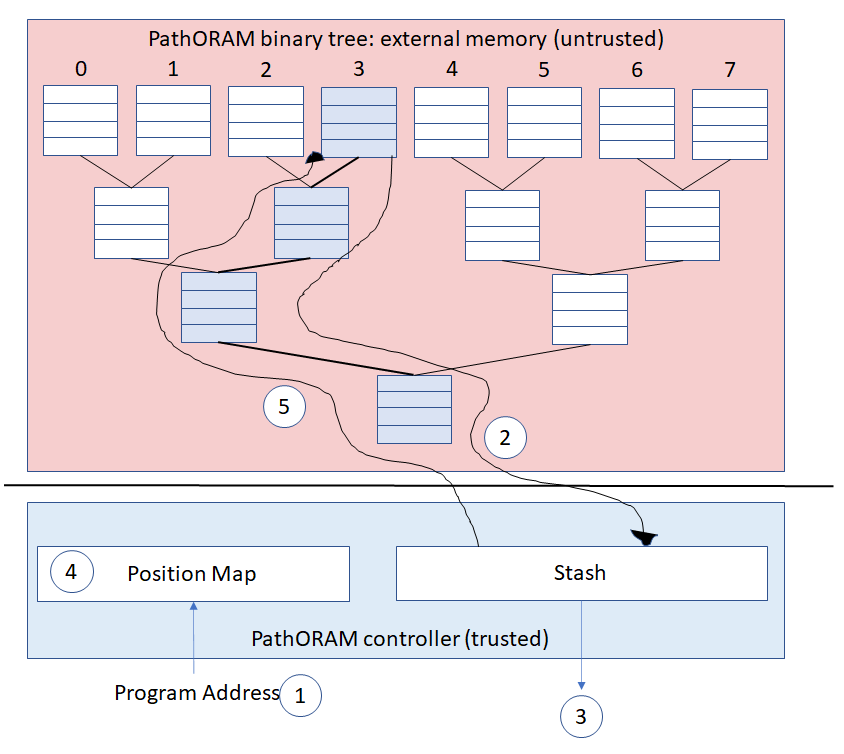}\\
  \caption{An example of PathORAM with 4 levels and access to path 3.}
  \label{fig:pathoram-img}
  \vspace{-5mm}
\end{figure}

Figure \ref{fig:pathoram-img} illustrates the various steps involved to access data in PathORAM. We provide a quick overview here and refer the reader to the original paper for more details. The PathORAM controller shown in the figure is considered the trusted client that maintains the position map and the stash. The PathORAM binary tree structure is maintained on the server side. Memory access on the server side are unprotected and must be protected through the ORAM strategy. The following is the sequence of steps performed in PathORAM. 

(1) The client generates a request to a data address. If the block is already in the stash, it is immediately provided. Otherwise, the controller looks up the position map with the block's address to identify the path where the requested data block is present in the binary storage. Let us assume that the path name is $S$. For simplicity, the path name in our illustration is simply the number associated with the leaf node.

(2) 
This path name $S$ is sent to the server and all the buckets along the path $S$, from the root node to the leaf node, are sent to the client. The client will add those blocks to the stash upon receipt. Some blocks within a node could be dummy blocks with no data, and the client may drop any block that is a dummy block from stash storage.

(3) The client then performs whatever operation is necessary on the requested data block.

(4) The controller then assigns a new path $S'$ for this data block. It then updates the path associated with the data block to $S'$ in the position map. This new path is chosen uniformly from all the leaf nodes.

(5) Finally, the controller writes the blocks in the stash that can be placed along the same path which was read, i.e., path $S$. Any blocks in the stash that cannot be placed in path $S$ are left in the stash. Also, any empty blocks along path $S$ are stored as dummy blocks.

The randomization of paths after every read/write is the foundation for PathORAM's ability to obfuscate address accesses, as the adversary sees paths are uniformly accessed on each data access. Both metadata management and bandwidth contention create significant access delays.

\subsection{PrORAM}
While PathORAM provides obliviousness, every access requires reading all the buckets along a path and metadata management, thereby causing a substantial bandwidth increase. PrORAM is a further enhancement over PathORAM using superblocks which were first introduced in ~\cite{superblock-pathOram}. A superblock is a combination of multiple data blocks with the condition that all the data blocks in the superblock are assigned the same path. A simple superblock e.g. is shown in figure~\ref{fig:superblock}. PrORAM takes advantage of past data locality and forms superblocks for consecutive data blocks. Two different approaches are presented in PrORAM:

\begin{figure}
\centering
  \includegraphics[width=8cm]{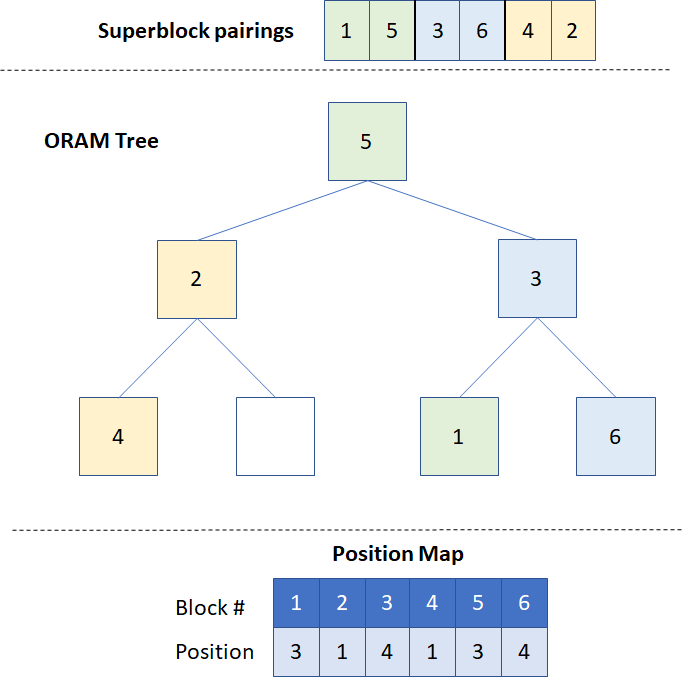}\\
  \caption{An example of superblock with superblock size of 2. All the data blocks in the superblock are given the same path. 1 and 5 share the same path, 2 and 4 share the same path, and 3 and 6 share the same path.}
  \label{fig:superblock}
  \vspace{-5mm}
\end{figure}

\textbf{Static Superblocks:} Static superblocks are formed by treating $n$ consecutive data blocks as one superblock. This approach is based on the premise that spatial locality exists across nearby blocks. 

\textbf{Dynamic Superblocks:} 
This approach utilizes a spatial locality counter. When two consecutive data blocks are accessed together, the spatial locality counter is increased. Once the spatial locality counter goes above a certain threshold, the data blocks are grouped into a superblock. If the blocks within a superblock are not accessed together, the counter value is decreased. Once the value goes below the threshold, the superblock is broken back into individual data blocks. By placing blocks that are likely to be accessed together, as predicted by the history-based locality counter, on a single path, the intuition is that more useful blocks can be fetched from each path.

\subsection{Stash Management}
The stash is small storage on the client side, used to buffer the data when a particular path is read from the binary tree storage. When the data block has been accessed, the path associated with the data block is changed, and the stash is written back along the same path which was read. This approach forces some data blocks to be left in the stash if overlapping nodes between their assigned path and accessed path have no space left. Excessive occurrence of this phenomenon can cause the stash to overflow. Background eviction has been used in PathORAM and PrORAM to reduce the stash size. Background eviction means that dummy reads are performed to random paths once the stash occupancy goes above a certain threshold. Dummy reads do not change any blocks' associated path so that they can reduce the number of blocks in the stash.

One way to reduce stash overflow is to leave more dummy blocks along each path, so there is a higher probability of finding an empty block in a path. For instance, PathORAM and PrORAM implementations suggest a tree structure with bucket size Z to have Z-1 dummy blocks in every node, but the space overhead of such an approach is large.

\section{Threat Model}
\label{sec:thread-model}
Before we describe the LAORAM, we first present the system setting and our threat model. 

\textbf{System setting:} 
Our system setting for embedding table training consists of three components: the server\_storage, the trainer\_GPU and the preprocessor. A CPU server stores the embedding tables in its DRAM. The server\_storage is equivalent to ORAM storage in PathORAM and is considered insecure. Hence, any memory address requests generated from server\_storage to its external memory interface is vulnerable.  But the content of the memory itself is considered encrypted and hence secure. In LAORAM the server\_storage is the CPU DRAM that is plentiful and cheaper than GPU memory. 

Training models such as recommender systems and NLP models that train the embedding tables on a CPU are quite slow.  Hence, the actual training process is performed on a \textbf{trainer\_GPU}. The GPU itself could be attached to a host CPU that stores the embedding tables. Alternately, the training GPU could be a remote GPU that can be reached over the network from the CPU. The training GPU has a limited amount of VRAM where it may cache the embedding table entries needed for an upcoming training batches. State of the art GPUs currently use high-bandwidth memory (HBM) as VRAM, which is integrated into the GPU package. Hence, when the GPU makes a request to access a cached embedding entry the address request to HBM stays within the package and is invisible to the adversary. 

While the above system setting is the default configuration in our experiments, we note that our approach can be easily adopted to other GPUs without HBM. However, in that case we need to cache the embedding entries within the GPU's on-chip shared memory, which can be easily done using existing memory allocation APIs in CUDA by tagging the cached embedding entries as \textit{\_\_shared\_\_}. Note that the size of the cached embedding entries is quite small since it only needs to store entries associated with at least one training batch, which is usually in the order of several hundreds of KB.  

The preprocessor is a special application thread in LAORAM that could run on the GPU client or any other trusted compute base. It preprocesses the training sample data and generates a list of embedding indices that may be used in the upcoming training batches. Communication  between all three components can be encrypted and hence invisible to adversary. 

\begin{figure}
\centering
  \includegraphics[width=8cm]{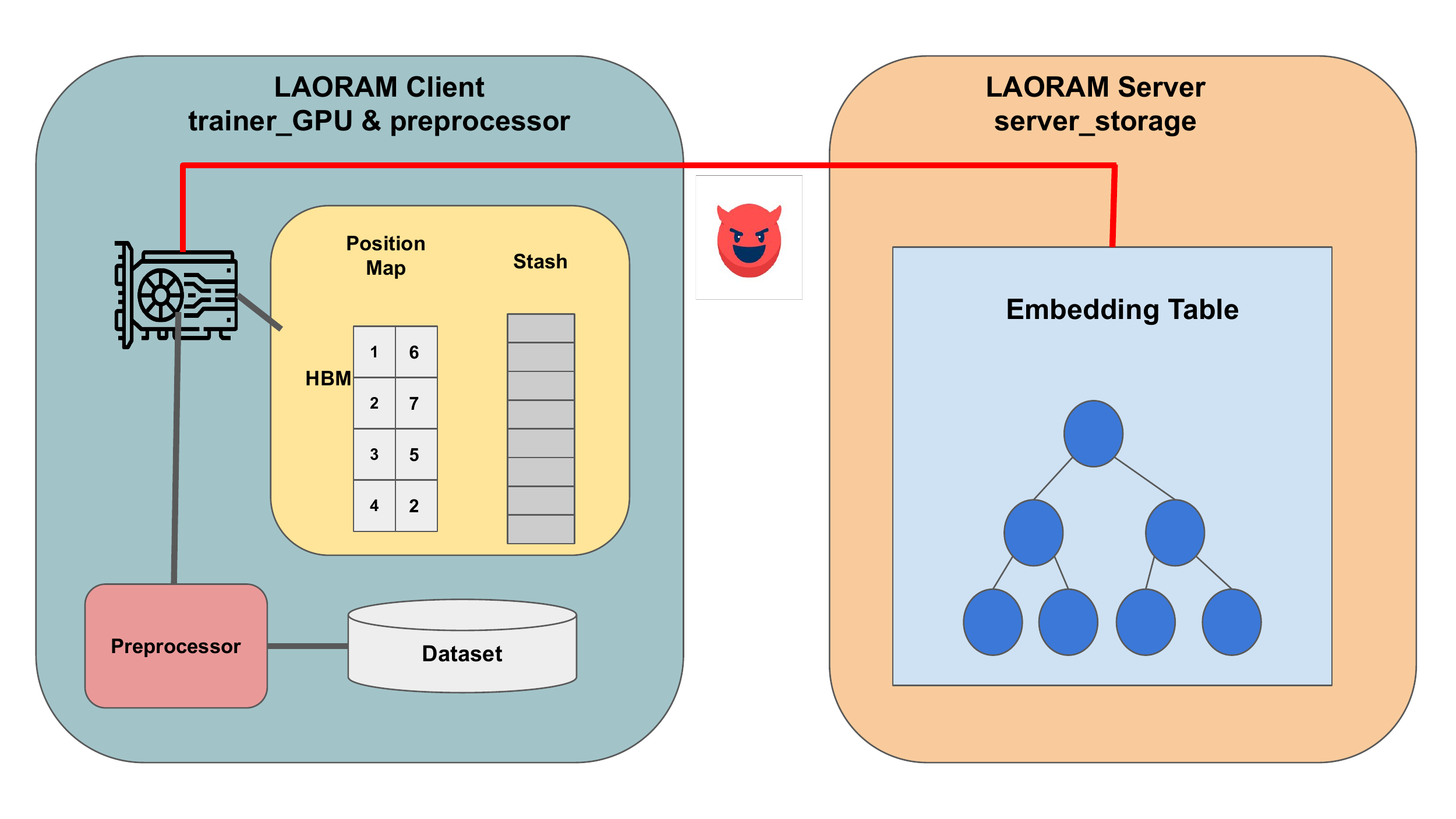}\\
  \caption{LAORAM Architecture Overview, black solid lines are secure communication channels and the red solid line is an insecure communication channel.}
  \label{fig:LAORAM-arch}
  \vspace{-5mm}
\end{figure}

\textbf{Threat Model:} 
LAORAM focuses on two types of adversaries. An adversary who can gain access to the CPU server and GPU and can probe the external buses to see the address patterns. Second, the adversary could be an honest-but-curious supervisor. LAORAM's goal is to prevent these two types of adversaries from obtaining the memory access pattern to the server\_storage where embedding tables are stored (see the red line in the figure~\ref{fig:LAORAM-arch}), and those access patterns directly reveal training input data. This implies that in our system setting the access address to an embedding entry in the CPU DRAM is visible to an adversary.  
Trainer\_GPUs acts as a client device in a traditional ORAM setting. Apart from caching a subset of the embedding entries needed for training batches, the GPU stores the stash and position map in their HBM VRAM. As explained earlier, accesses to the HBM are invisible to an adversary even if they can physically access the device. Even if there is no on-package memory,  the cache, stash and position maps can be allocated within the on-chip shared memory. 

The Trainer\_GPU may reside in the same host as the CPU server\_storage, and may be controlled by the same curious OS that controls the CPU. In this case, GPU computations can be protected from a curious OS using  GPU enclave techniques~\cite{hix,graviton}, or confidential computing techniques in the new generations of GPUs~\cite{confidential-compute-gpu}.  Those secure GPU techniques, like CPU enclaves, can refuse unlawful queries from host OS. Moreover, GPU enclaves can have a single user enclave thread running inside a GPU further disabling the host OS from establishing side channel attacks to discern data access pattern. 




\section{LAORAM Algorithm and System Design}
\label{sec:laoram}
ML model training has the unique advantage where the future memory access patterns to the embedding tables are known beforehand because the training samples that will be used for the next several batches can be analyzed beforehand.  LAORAM exploits this knowledge  in  forming superblocks. As explained earlier, superblocks are multiple data blocks grouped together such that all the data blocks in the superblocks are assigned the same path. 

LAORAM merges embedding table entries that will be accessed in the upcoming training batches into superblocks. Reads and writes in LAORAM happen at the granularity of superblocks.  If the superblock size is four, a single access to the superblock means accessing all four of them. Once the superblock is accessed, the path of all four data blocks is changed independently based on their future locality. Hence, the fact that the group of blocks are formed into a superblock during one access has no bearing on how those blocks are paired in future. For instance, if all four blocks are going to be accessed separately with other data blocks in the future their future paths would be different. If only two of the four blocks are going to be accessed together in the future then only those two blocks are placed together in a superblock, the other two would be in separate superblocks.

We now describe the overall architecture and the detailed algorithmic design of the LAORAM. The approach has two main operations: The first one is the preprocessing which reads a batch of upcoming training samples and identifies embedding table entries that are going to be accessed together, and determine future paths for those entries. The second operation is the embedding table training using information provided by preprocessing.

\subsection{LAORAM Architecture}
\label{sec::LAORAM:arch}
Figure~\ref{fig:LAORAM-arch} provides an architecture overview of LAORAM. Like traditional ORAM, LAORAM has two major components: a LAORAM server and a LAORAM client.

\subsubsection{LAORAM Server(Server\_storage):} 
The LAORAM server, referred to as server\_storage, is the CPU DRAM storing embedding table entries in a binary tree ORAM structure. It is equivalent to a traditional ORAM server. The LAORAM server is responsible for providing the embedding table entries requested by the GPU trainer. Accesses to the CPU DRAM are considered insecure and an adversary can observe the address patterns to the server\_storage. Hence, all accesses sent to the CPU DRAM on the LAORAM server must be obfuscated using the path oblivious scheme of LAORAM.

\subsubsection{LAORAM Client:}
\textbf{The trainer\_GPU} uses its VRAM to store the position map and stash memory, along with the cached embedding entries that are trained on the GPU, and it performs gradient descent operations on the cached embedding table entries to converge to the final ML goal. The position map and stash serve their traditional roles as in PathORAM. In particular, the position map stores the path where each memory block is stored in the ORAM tree. The stash stores all the data blocks that cannot be evicted to the server\_storage. The trainer\_GPU will use those data structures and metadata received from the preprocessor to ask the host CPU to fetch embedding table entries it needs at a given time and store them in the VRAM. In LAORAM, instead of requesting the actual embedding table entries, the trainer\_GPU will request the corresponding paths associated with those embedding table entries to prevent leakage. 

\textbf{The preprocessor}, a special LAORAM client component not present in the traditional ORAM designs, is responsible for preprocessing upcoming training samples and extracting embedding table entries that are going to be accessed together to form superblocks. It will also assign a random path to each newly formed superblock. Information about superblock formations and the random path associated with each superblock are communicated to the trainer\_GPU. The trainer\_GPU will assign the paths to data blocks based on the information received from the preprocessor and updates its position map. The GPU then issues read request to all the paths associated with the embedding entries in the upcoming training batch and caches them locally before starting the batch training process.

\subsection{Preprocessing Algorithm}
\subsubsection{Algorithm overview}
As mentioned in the previous section, the preprocessor is responsible for analyzing future training batches and identifying the list of entries that will be accessed together to form superblocks. This subsection describes the preprocessing algorithm. Preprocessing consists of two steps: 1) Dataset scan, and 2) Superblock path generation.

\subsubsection{Dataset scan}
The preprocessing algorithm needs the superblock size $S$ as the input. The preprocessing node looks at the upcoming training batches and places the next $S$ entries into a superblock bin. It continues this binning process to create as many bins as it can preprocess while staying within the compute and memory limitation of the preprocessing node. For instance, the preprocessing node may scan an entire epoch of training batches if it has sufficient memory capacity to create and store all the bins necessary to hold the entries within the epoch. 

\subsubsection{Superblock path generation}
In this step, the preprocessor will pick a random path for every superblock bin formed during the scanning step. Paths are chosen from a uniform distribution of $U(1,L)$ where $L$ is the number of leaves in the PathORAM. That is to say, a path for each superblock bin is chosen uniformly from among one of the leaf nodes. Thus each superblock bin is now assigned a path number. After each superblock bin gets its path, the preprocessor will produce a superblock to future path number mapping. This (superblock, {future path numbers}) metadata, where path numbers in the future path number set are the future path for data blocks in the superblock, is then securely transmitted to the trainer\_GPU. The trainer\_GPU stores this metadata to assign predetermined future paths to data blocks when accessed.

\section{Fat Tree for Stash Efficiency}
\label{sec:fattree}
\textbf{Increased Stash Usage using Superblocks:}
One potential challenge with using superblock designs in ORAM is the stash growth. With superblocks multiple blocks are going to be placed in the same path. Hence, when the trainer\_GPU needs to write the data back to the server\_storage it has to find sufficient space along a given path to accommodate all the blocks within a superblock. When some of the blocks in the superblock cannot be accommodated in the given path those blocks must be temporarily held in the client's stash memory. By forcing a single path assignment to all the blocks within a superblock, the probability that a block must be held in stash increases. 
When the stash is full, trainer\_GPU can issue dummy evictions (read random paths without accessing any data block) to drain the stash. These dummy reads will degrade performance but are the only way traditionally ORAMs can reduce stash usage.

\textbf{Naive Increase Bucket Size:}
One solution to this problem is to increase the bucket size of the nodes in the tree, which will reduce the stash size. However, such a naive solution does not efficiently utilize added memory spaces.

\begin{figure}[h]
\centering
  \includegraphics[width=8cm]{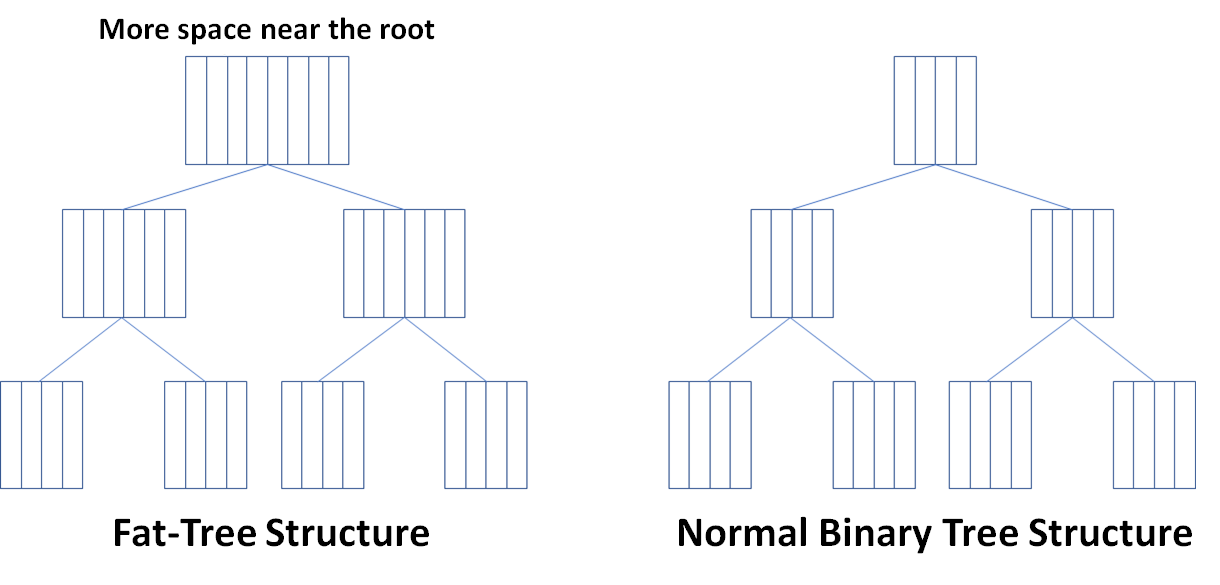}\\
  \caption{Fat-tree vs normal binary tree. Fat-tree has more buckets near the root and hence has lower chances for path to be left in stash.}
  \label{fig:fattree}
  \vspace{-3mm}
\end{figure}

\textbf{Key Observation about Buckets in Different Levels:}
One key observation we have made is that the probability of a data block being written in a particular level goes down as we go to the leaf node in the tree. In particular, the probability of a block being written back to root is 0.5, the probability of a block being written to a level 1 node is 0.25, and so on. With this insight, we propose to use a variable bucket size with wider buckets near the root and narrower buckets near the leaf. This structure draws parallels to  the fat-tree network topology~\cite{fat}, as shown in figure \ref{fig:fattree}. For example, when the original bucket size is 5, the bucket size at the root will be 10 and it decays to a bucket size of 5 near the leaf.

\textbf{Fat-tree organization:}
In our fat-tree implementation, bucket sizes are tied to the leaf bucket size. If the leaf bucket size is $x$ then the bucket size at the root node is $2x$. For example, if the leaf node bucket size is 5 and the number of levels in the tree structure is 6, the root (level 0) will have a bucket size of 10, level 1 will have a bucket size of 9, and so on until the leaf level has a bucket size of 5.
Ideally, the bucket size should grow exponentially as the level gets closer to the root to provide the same probability of finding an empty space at each level. However, exponential growth is not practical in ORAM tree organization due to huge overheads at the root.
Thus, we choose linear growth. Despite the fact that our bucket growth is not exponential, our fat-tree design is able to improve the utilization of memory spaces at each tree level to accommodate data blocks sent from the stash without causing undue stash growth.  

Compared with increasing bucket size for the entire tree uniformly, our fat-tree structure can trigger fewer evictions while slightly increasing the memory requirement for wider buckets near the root. Wider buckets near the root give more opportunity for data blocks to be written back to the server\_storage, and hence the fat-tree structure uses  memory space more effectively. As we show later, the fat-tree triggers $12.4\%$ fewer evictions while using $16.6\%$ less  memory space compared to doubling the bucket size uniformly.

\section{Security Analysis}
\label{sec:security}
LAORAM consists of an LAORAM server i.e. server\_storage and an LAORAM client (trainer\_GPU \& preprocessor). Each subsection below presents the security analysis of path obliviousness of those components.
\subsection{ORAM Server (server\_storage) Access Patterns}
This section will discuss the obliviousness of memory access patterns to the LAORAM server (server\_storage) where embedding tables are stored. Server\_storage will read data from CPU DRAM in path granularity. These path requests originate from the trainer\_GPU. These path granularity accesses are oblivious because memory accesses to the superblocks are identical to those of PathORAM, and PathORAM access patterns are oblivious. Below we prove that using superblocks does not reveal any additional information over using memory blocks.  

\textbf{Obliviousness of Superblocks:}
When a superblock is accessed, the path the superblock is assigned to will be fetched.
Although multiple data blocks in a path are accessed, every block in the superblock will be assigned to a new uniformly chosen path. Uniformity of new path assignment is proved using total probability\cite{stat-book}. Let $N$ be the total number of paths in the binary tree, $p$ be any path in the binary tree, and $b_i$ be superblock $\#i$. Additionally, let's define $d\bar{\in} b_i$ as a data block $d$ being firstly accessed in superblock $\#i$ and $p\in b_i$ as path $p$ assigned to the superblock $\#i$. Then, we can prove the uniformity:
\begin{align}
    &P\{NEXT\_PATH(d) = p\} \nonumber \\
    \label{step1-tp}
    &= \sum_{i=1}^{\infty}  P\{NEXT\_PATH(d) = p |d\bar{\in} b_i\}\cdot P\{d\bar{\in} b_i\} \\
    \label{step2-def}
    &= \sum_{i=1}^{\infty} P\{ p\in b_i |d\bar{\in} b_i \} \cdot P\{d\bar{\in} b_i\} \\
    \label{step3-ind}
    &= \sum_{i=1}^{\infty} P\{ p\in b_i\} \cdot P\{d\bar{\in} b_i\} = \sum_{i=1}^{\infty}  \frac{1}{N} \cdot P\{d\bar{\in} b_i\} \\
    &= \frac{1}{N} \cdot \sum_{i=1}^{\infty}  P\{d\bar{\in} b_i\} =\frac{1}{N}\cdot 1 = \frac{1}{N}
\end{align}
Equation \eqref{step1-tp} holds due to total probability. Equation \eqref{step2-def} holds due to the definition of the conditional probability in \eqref{step1-tp}. Equation \eqref{step3-ind} holds because path assignment is independent of data blocks in the superblocks. The rest of proof just follows arithmetic manipulations.

After path reassignments, the original path the superblock associated with will the written back using data blocks currently in the stash. Thus, memory access patterns of superblocks are identical to those of the original PathORAM. Adversaries can only observe that random paths are being accessed every time embedding table look ups take place \cite{proram}.

\textbf{Original PathORAM:}
The security of the original PathORAM is guaranteed by statistical independence of the old path a data block belongs to and the future path the data block is going to be assigned to. Every new path is chosen from an independent uniform distribution. Consequently, when a data block is being accessed the second time, a new independent path will be fetched. Adversaries will have a negligible probability to track what data blocks are being accessed.

Since the usage of superblocks will generate the same memory access behaviors as the original PathORAM and the original PathORAM is oblivious, LAORAM accesses to ORAM server memory are also oblivious.

\subsection{Trainer\_GPU}
The Trainer\_GPU acts as a client in our setting, which is considered trusted in ORAM designs. As we mentioned earlier in section~\ref{sec:thread-model}, there exists techniques to protect GPUs against a curious OS which are sufficient for our threat model.

\subsection{Preprocessor}
In LAORAM, the preprocessor reads the training samples and is considered part of the secure client. The memory accesses generated during preprocessing only access training samples which are encrypted data. Note that the training sample address is not a security concern. It is the embedding entries that are listed within the training samples that must be protected. Hence, the preprocessor can process the samples to extract these embedding entry values within the the ORAM client. But these entry values are not used to access any memory within the preprocessor. Thus, we consider the preprocessor is secure.

\section{EXPERIMENTAL SETUP}
\label{sec:evaluation}
\subsection{Methodology \& Terminology}
We implemented LAORAM as a self-contained memory access engine which has the ORAM server and client architecture functionality implemented across a GPU (ORAM client i.e. trainer\_GPU), and CPU with DDR4 memory (server\_storage). This infrastructure can be integrated within any embedding table training system.  Our system takes a series of memory block addresses generated by a training data preprocessor as inputs and creates superblocks and assigns paths as per the algorithms described earlier. We demonstrate performance improvements of LAORAM with and without fat-tree against the baseline PathORAM.

\subsection{Models and Datasets}
We use the following datasets and models  to quantify the benefits of LAORAM: Permutation Dataset, Gaussian Dataset. The \textbf{Permutation dataset} randomly generates an address in the range $0-N$  where none of the addresses are repeated until all the addresses are accessed at least once. The \textbf{Gaussian dataset} generates an address stream sampled from Gaussian distribution. In addition to the two datasets we also use two ML  models that train embedding tables and their associated datasets to evaluate LAORAM. We use the DLRM recommendation model~\cite{dlrm} that trains large embedding tables using the Criteo AI Labs Ad Kaggle Dataset (Kaggle). The \textbf{Kaggle dataset} is released by Criteo AI labs containing real user data and is provided by DLRM as one of its representative recommendation model training dataset. Finally, we use XLMR~\cite{xlmr} a well known natural language processing model that trains embedding tables using the Cross-Lingual NLI Corpus (XNLI) dataset. The \textbf{XNLI dataset} contains multiple data points in multiply languages, and it is used as a key metric for NLP models~\cite{xlm}. 

The permutation dataset will produce LAORAM's worst-case performance because it will put the biggest pressure on the stash size (proven in the original PathORAM paper~\cite{path-oram}). Bigger pressure on stash size results in more dummy reads, leading to the worst case LAORAM performance. The DLRM and XLMR models and their associated datasets show the LAORAM performance under practical application setting. 

In our results (figure~\ref{fig:speedups}), we label results that LAORAM with fat-tree structure as``Fat'', and results that use LAORAM only without fat-tree as ``Normal''. We also use $/S\#$ notation to indicate the superblock size. For example, "Fat$/S2$" means that the given tree uses a fat-tree structure and the superblock size is $2$. We used PathORAM~\cite{path-oram} as the baseline, which can be treated as an ORAM design with superblock size of just one.  Note that using PrORAM~\cite{proram} with highly random embedding table accesses (see Figure~\ref{fig:random-access}) degrades its performance to be the same as PathORAM, even after ignoring the superblock tracking and formation overheads. Hence our results use PathORAM with a superblock size of 1 as the baseline.

\subsection{Embedding Table Configurations}
For both Permutation and Gaussian, we have chosen the total number of embedding entries to be 8M and 16M. For the DLRM embedding tables the largest table from Kaggle has $10131227$ entries. Each embedding table entry is $128$ bytes. For XLMR model the embedding table has $262144$ entries and each entry is $4k$ bytes. The memory required to hold those embedding tables is shown in the table~\ref{table:sizes}. The default bucket size for PathORAM and LAORAM is $4$. Note that while one may consider these chosen memory sizes to fit within the high end GPU HBMs, we chose these sizes to stay within our experimental infrastructure limits on GPU and CPU physical memories. It has already been demonstrated that the size of the embedding tables for recommendation systems in production environments can exceed 100s of GB~\cite{aibox-one-node-training, mudigere2021high}. Similarly, based on current trend, with NLP models supporting more languages, we expect the size of embedding tables in NLP to grow as well.

\begin{table}[h]
\centering
 \begin{tabular}{||c |c | c | c | c ||} 
 \hline
  & Insecure & PathORAM & LAORAM &  FAT  \\
 
 \hline\hline
  8M & 1GB & 8GB & 8GB & 10GB \\ 
 \hline
  16M & 2GB & 16GB & 16GB & 24GB  \\ 
 \hline
  Kaggle & 1.2GB & 16GB & 16GB & 20.3GB  \\ 
 \hline
  XNLI & 1GB & 16GB & 16GB & 20.5GB  \\ 
 \hline\hline

\end{tabular}
\caption{Embedding table memory requirement}
\label{table:sizes}
\vspace{-5mm}
\end{table}

\subsubsection{System configuration}
We conducted our experiments on an experiment platform consisting of an Intel  Xeon  E-2174G CPU as the ORAM server with 64GB DDDR4 memory to store the embedding tables. And a RTX 1080 Ti GPU based system is used as an ORAM client. 

\begin{figure*}[htbp]
  \centering
  \begin{subfigure}[tb]{0.3\linewidth}
    \includegraphics[width=\linewidth]{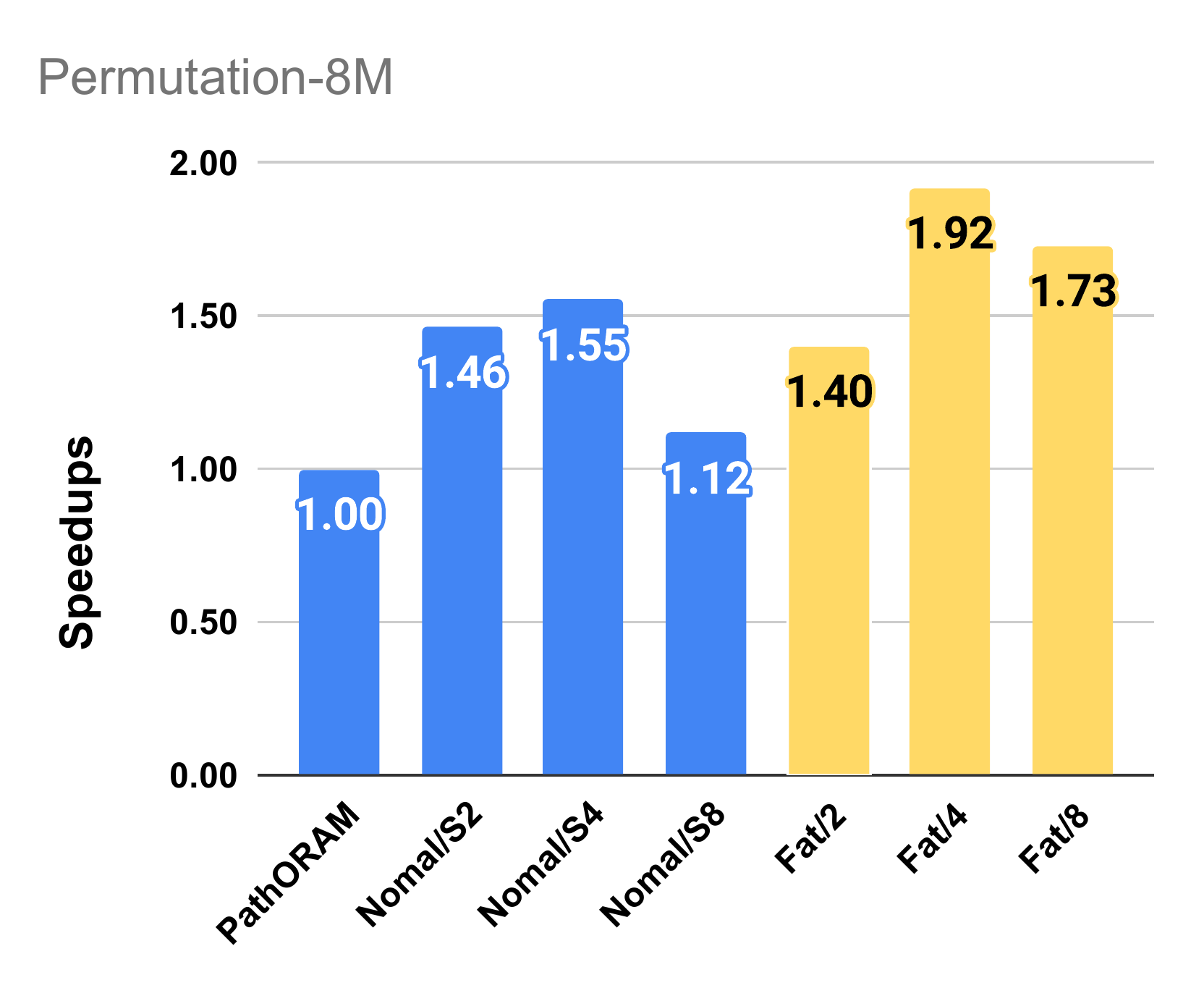}
    \caption{8M permutation}
    \label{fig:perm-speedup1}
  \end{subfigure}
  \begin{subfigure}[tb]{0.3\linewidth}
    \includegraphics[width=\linewidth]{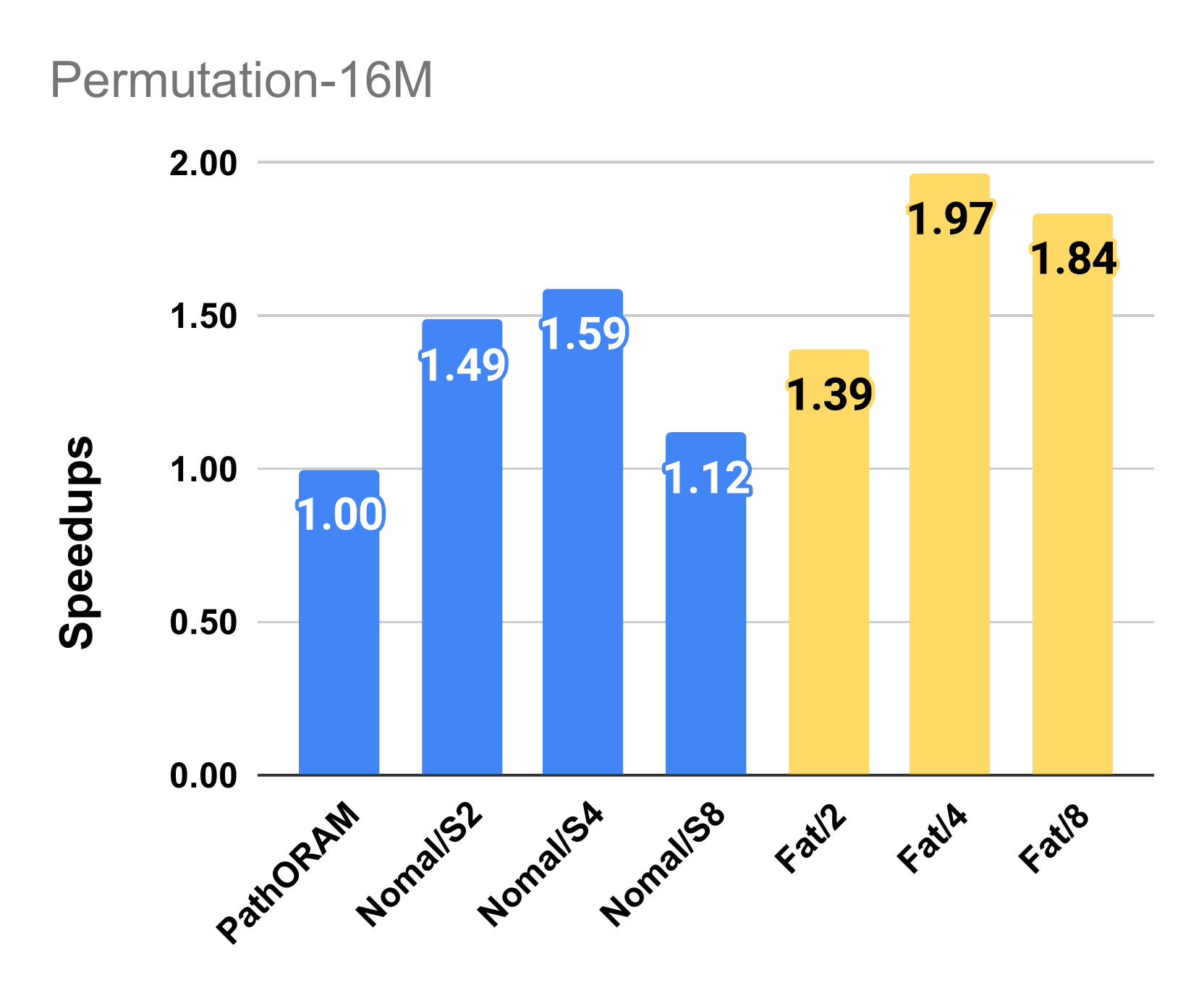}
    \caption{16M permutation}
    \label{fig:perm-speedup2}
  \end{subfigure}
    \begin{subfigure}[tb]{0.3\linewidth}
    \includegraphics[width=\linewidth]{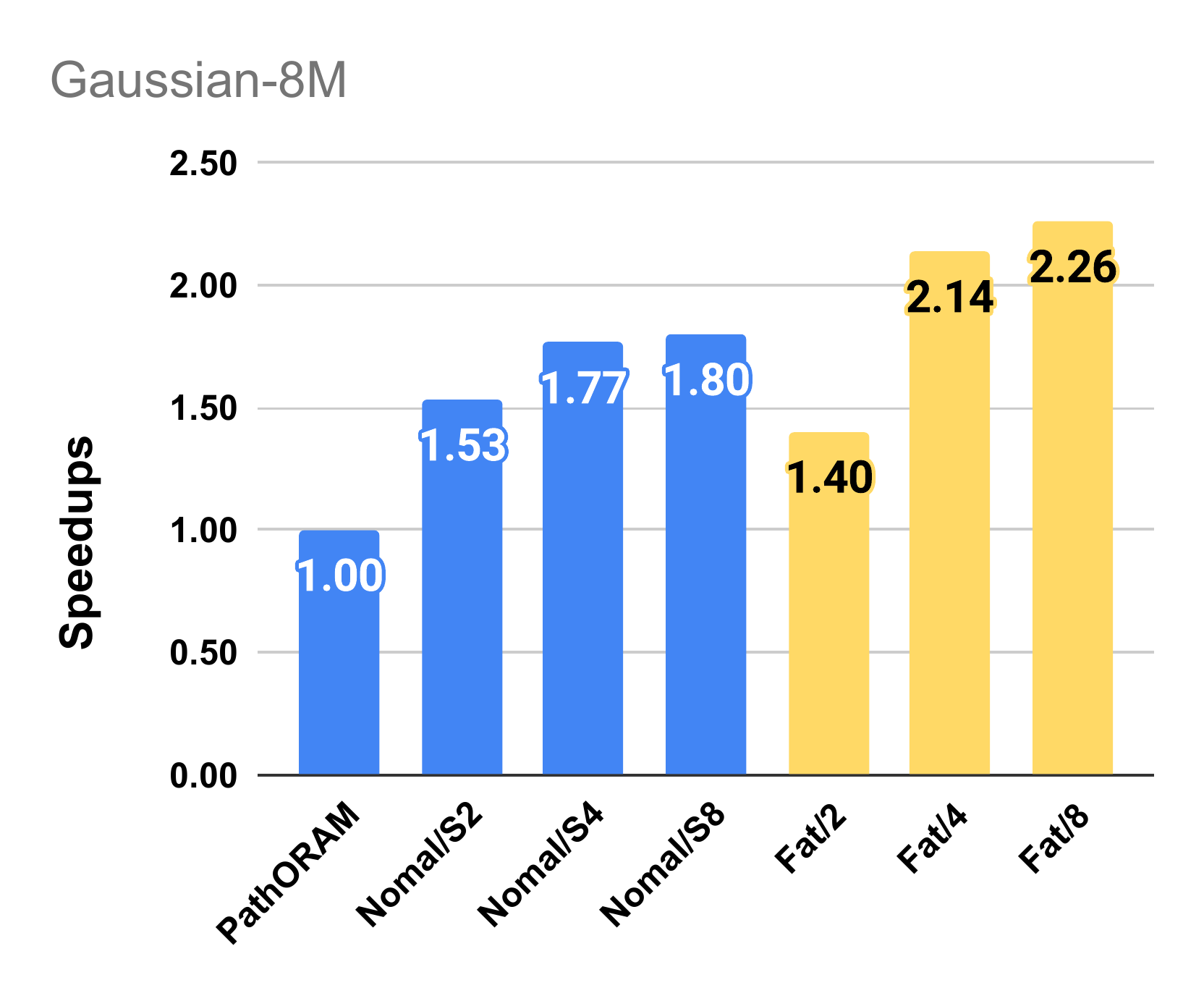}
    \caption{8M Gaussian}
    \label{fig:gaus-speedup1}
  \end{subfigure}
  \label{fig:worst-speedups}
  \begin{subfigure}[tb]{0.30\linewidth}
    \includegraphics[width=\linewidth]{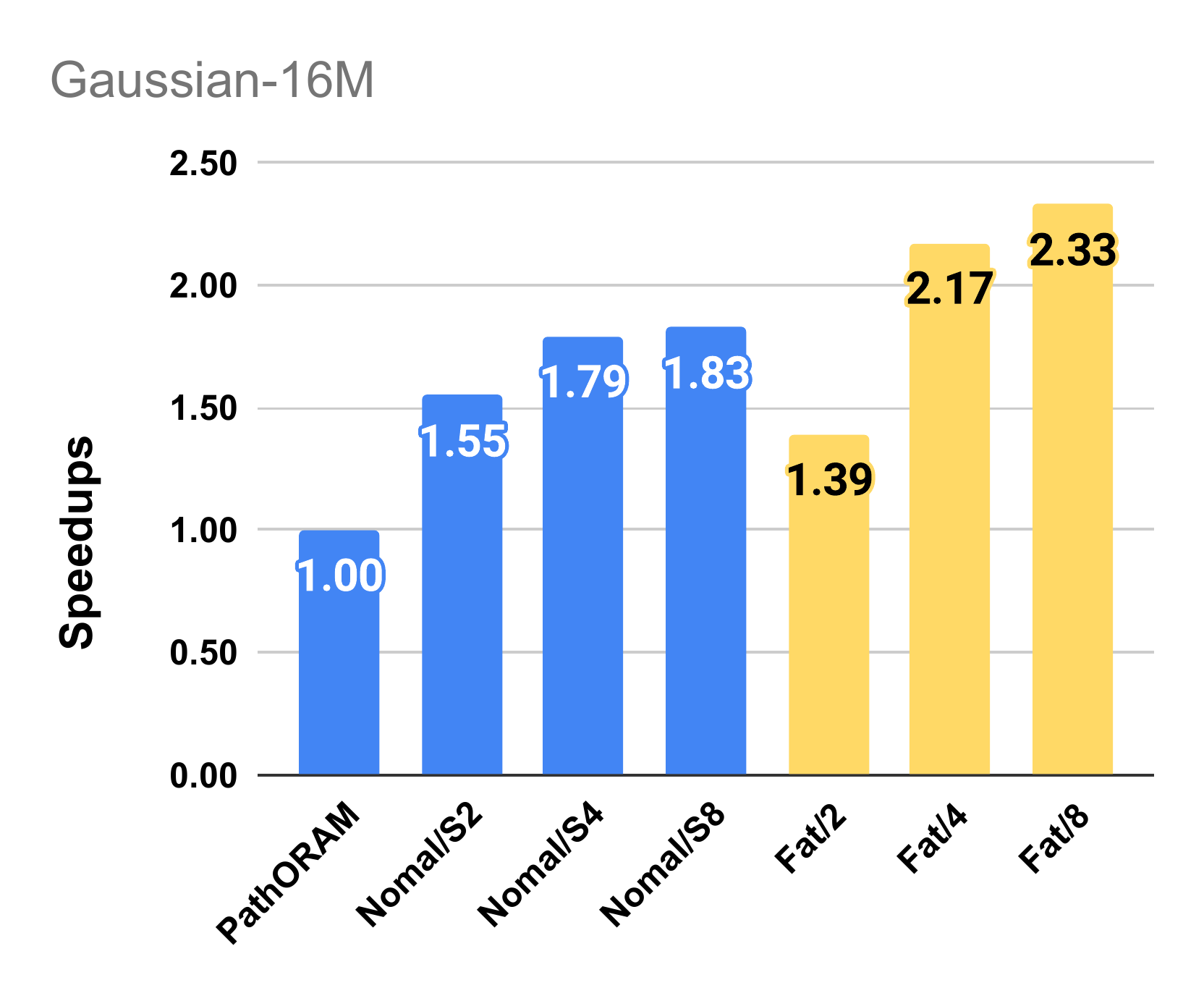}
    \caption{16M Gaussian}
    \label{fig:gaus-speedup2}
  \end{subfigure}
  \begin{subfigure}[tb]{0.30\linewidth}
    \includegraphics[width=\linewidth]{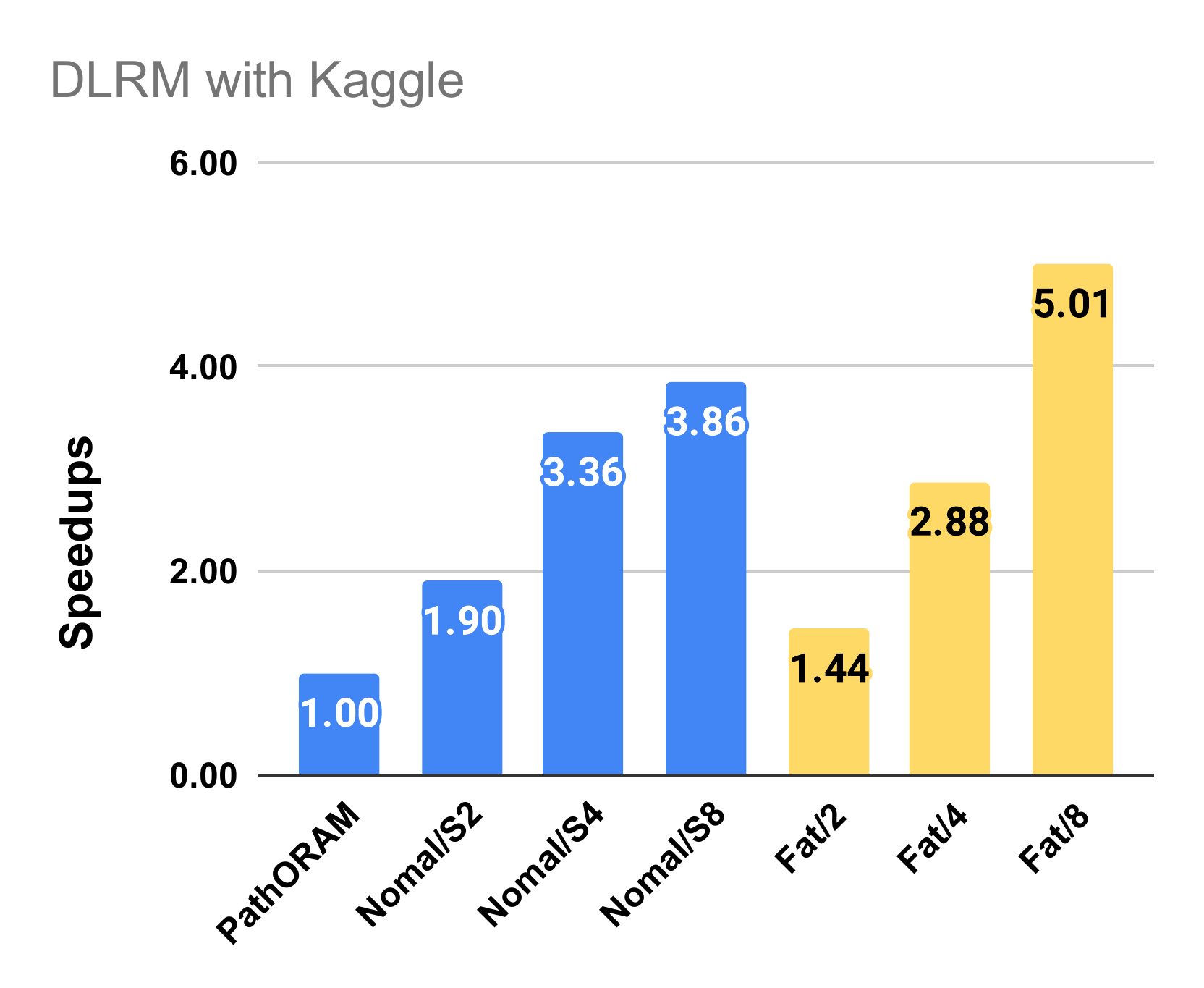}
    \caption{DLRM with Kaggle}
    \label{fig:kaggle}
  \end{subfigure}
  \begin{subfigure}[tb]{0.30\linewidth}
    \includegraphics[width=\linewidth]{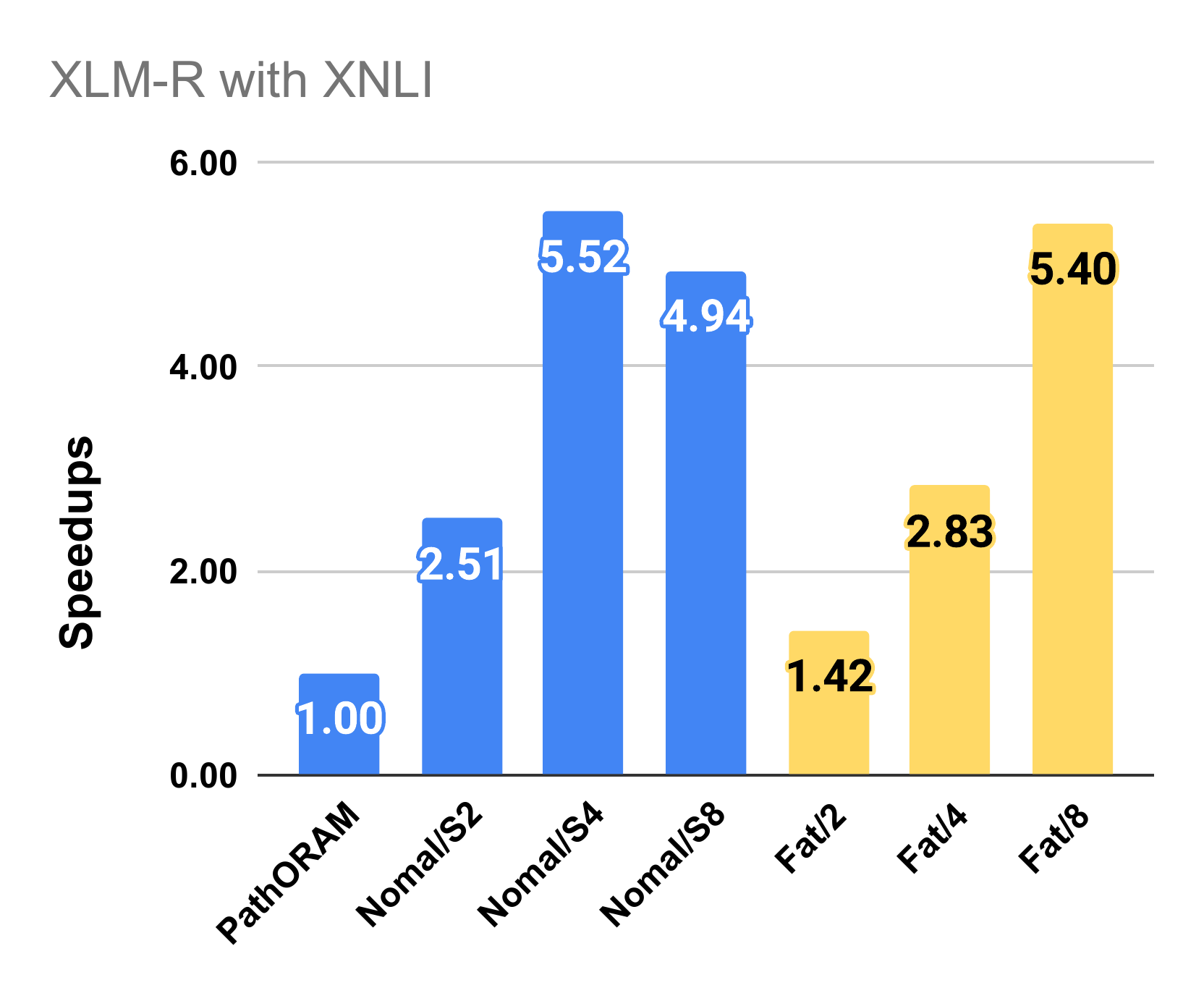}
    \caption{XLM-R with XNLI}
    \label{fig:xnli}
  \end{subfigure}
  \caption{LAORAM Speedups.}
  \label{fig:speedups}
\end{figure*}

\section{Experimental Results}
\subsection{Preprocessing timing}
Preprocessing and accessing data are two pipeline stages in the 2-stage LAORAM pipeline. Once the preprocessing for the first several batches is complete, GPU can generate the LAORAM accesses and start the training process. The preprocessing can then run ahead of the GPU training process. In our experiments, preprocessing a sample to identify unique indices is orders of magnitude faster than training that sample on a GPU. 
Thus, the runtime shown in later sections does not include preprocessing since it is not on the critical training path and insignificantly small.

\subsection{LAORAM Access Latency Results}
\textbf{Permutation Dataset:} In this section, we evaluate the performance improvement of LAORAM over a PathORAM baseline for the worst-case permutation dataset. For these results, we measured the amount of time it took to access a given block from a client's computation viewpoint. The time includes sending the path to be fetched from the trainer\_GPU client to the server\_storage, the server\_storage accessing the tree and fetching all the blocks in the path, and the server\_storage sending that data back to the trainer\_GPU's stash.  As shown in figure \ref{fig:perm-speedup1} with 8M permutation data set, using LAORAM with normal tree structure and a superblock size of 2 and 4 gives us a speedup of 1.46X and 1.55X, respectively, over a path ORAM baseline. The speedup shows that forming superblocks using LAORAM's tracking of the upcoming embedding table indices increases performance. Nevertheless, when the superblock size increase to 8, the performance dips to 1.12X because a superblock size of 8 puts greater pressure on the stash and increases the number of dummy reads by $100\%$, compared to using superblock size of 4.  Although, when using permutation dataset, $fat/S8$ does not outperform $fat/S4$ due to pressure on the stash, in later results, $fat/S8$ can outperform $fat/S4$ when using real world data stream.

The last three bars in Figure \ref{fig:perm-speedup1} shows the performance of LAORAM with a fat-tree. As mentioned before, a fat-tree structure is most useful when there is stash contention and background evictions. With large superblocks, the contention grows, and the fat-tree is designed to handle large superblocks more efficiently. Hence, when using superblock size of 2, the fat-tree organization sees limited benefits, but with a superblock size of 4 and 8, the fat-tree implementation achieves substantial performance boost compared to the baseline and outperforms the normal tree implementation. We observe similar behavior for permutation data set size of 16M as shown in figure \ref{fig:perm-speedup2}. 

\textbf{ML models}   
This section discusses speedups for DLRM, XLMR models along with the Gaussian datasets.
Figure \ref{fig:gaus-speedup1}, \ref{fig:gaus-speedup2}, \ref{fig:kaggle} and \ref{fig:xnli} present results for 8M-Gaussian, 16M-Gaussian, DLRM with Kaggle datasets and XLMR with XNLI datasets. In the original PathORAM paper, probability of stash overflow is maximized when access patterns contain no duplicate addresses (which is emulated in our Permutation dataset). For  DLRM and XLMR, there are some duplicate addresses within a given window of  access stream. 

Note that in Figure~\ref{fig:random-access} even though the address stream is heavily randomized, there is a thin band of repeated accesses at the bottom of the figure. Such repetitive accesses, even when they are a small fraction, do improve the likelihood of placing that block back into the LAORAM path. Hence, the stash growth is reduced. This reduction in turn reduces the number of dummy reads needed later to reduce the stash size. 
Thus, LAORAM sees higher performance improvements due to this repetition of address accesses.

\subsection{Memory neutral comparison} 
Using fat-tree increases the storage requirement of the tree structure compared to normal tree with the same number of levels. The increase in storage requirement for the fat-tree compared to the normal binary tree for 8 million and 16 million embedding entries is  25\% and 50\%, respectively. 
To discount the benefit of this additional memory, we have also conducted  experiments where the normal tree's bucket size is increased such that the total size of the normal tree is at least as big as the fat-tree. In particular, we conducted an experiment where the normal tree uses a bucket size of 6, and a fat-tree's bucket size varies from 5 (leaf node) to a bucket size of 9 (root) for the same number of tree levels. In this organization the fat-tree uses $16.6\%$ less memory space than the normal tree. While we do not plot the results here due to space limitation, our results show that even with such a memory size handicap, the fat-tree generated $12.4\%$ fewer number of dummy reads. Increasing the bucket size of all nodes in the tree uniformly does not provide as much performance gain as the fat-tree design because the fat-tree uses memory spaces more effectively by allocating more space for nodes with higher probability to accommodate data blocks in the stash.

\subsection{Stash Usage}

\begin{figure}
\centering
  \includegraphics[width=8cm]{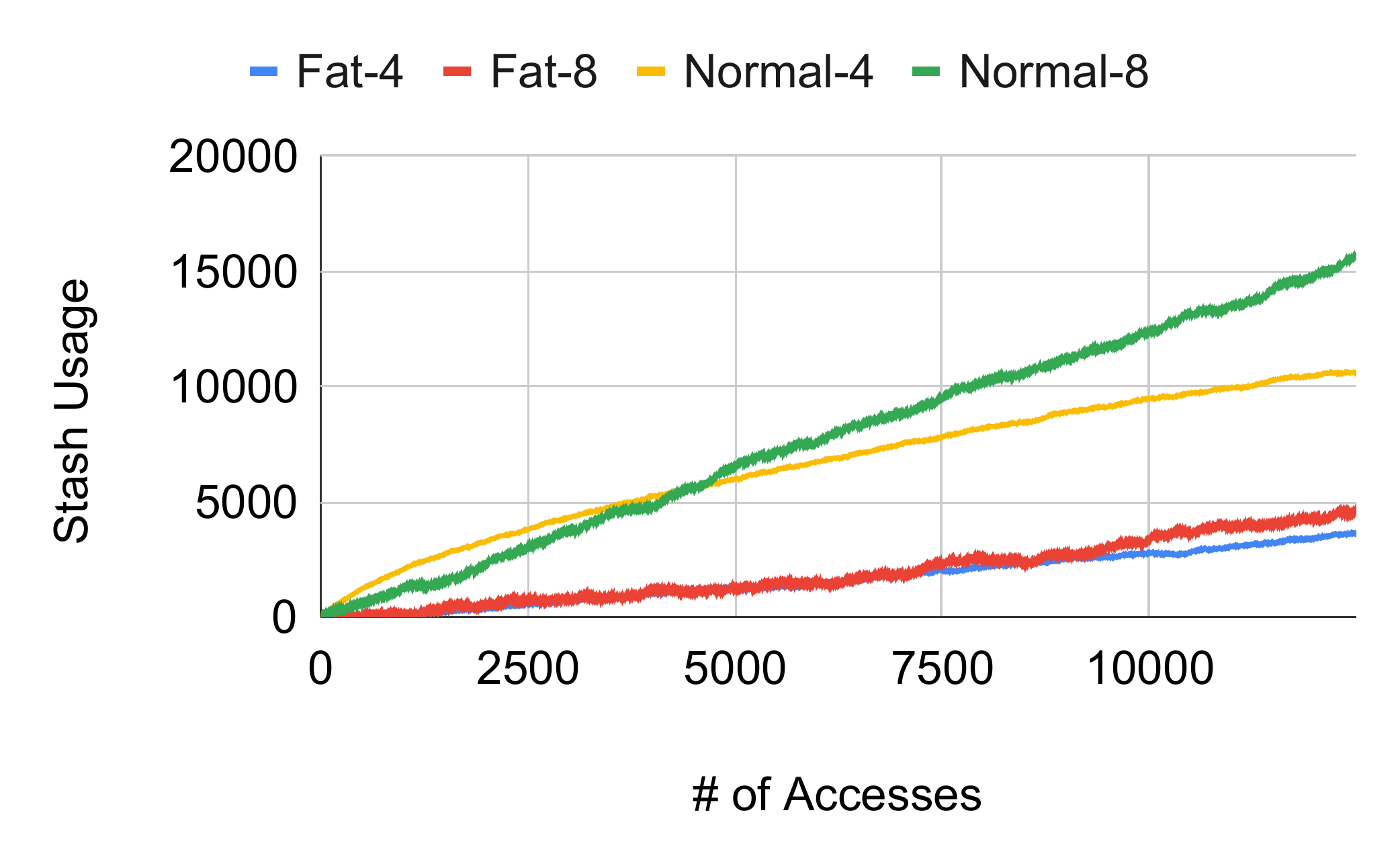}
\caption{Fat-tree vs normal binary tree. Fat-tree incurs less blocks in the stash when superblock size is large.}
  \label{fig:fat-tree}
  \vspace{-5mm}
\end{figure}

In figure \ref{fig:fat-tree}, we show the stash growth of LAORAM with and without fat-tree using 2 different configurations. The $fat/4$ and $normal/4$ configurations involve a superblock size of 4. We use a bucket size of 4 for normal tree and 8-to-4 for fat-tree. Fat-8 and Normal-8 configurations involve a bucket size of 8 and 8-to-16.  The graph shows that after around 12500 accesses, the stash for a $normal/4$ grows to around 10600 blocks, while the stash for $fat/4$ grows to about 3600 blocks.  After 12500 accesses, $normal/8$ has a stash size of 15500 blocks while $fat/8$ has a stash size of 4700. These results show that for larger block size the stash growth for fat-tree is slower than that of normal tree.

\subsection{Dummy Read Frequency} 
One way to drain the stash is to perform background evictions whenever the stash usage gets above a particular threshold. In this section, we measure the average number of dummy reads needed per access. In this experiment dummy, reads are triggered whenever the stash size grows above 500 entries, and a series of dummy reads are performed until the stash size reduces to 50 entries. 

Table \ref{table:dummyratio} shows the average dummy reads per access for both normal tree structure and fat-tree structure for different superblock sizes for Permutation, Gaussian, and DLRM. Permutation dataset presents the worst pressure on the stash for any PathORAM~\cite{path-oram}. For the Permutation dataset, for $normal/4$ and $normal/8$, the average number of dummy reads per access is $0.57$ and $1.19$. That for fat-tree is $0.14$ and $0,24$. When using the fat-tree structure, the total number of dummy reads is reduced by $60\%$ and $75\%$. A similar phenomenon can be observed for  other models. Usage of the fat-tree  reduces the number of dummy reads required by nearly 3 times.

\begin{table}[h]
\centering
\begin{tabular}[width=8cm]{|c|c|c|c|c|}
\hline
\textbf{Config} & \textbf{Permutation}& \textbf{Gaussian}  & \textbf{Kaggle}  & \textbf{XNLI} \\ \hline
\textbf{Fat/S8} & 0.35 & 0.24 & 0.025 & 0.009 \\ \hline
\textbf{Fat/S4} & 0.14 & 0.10 & 0 & 0\\ \hline
\textbf{Normal/S8} & 1.19 & 0.65 & 0.19 & 0.16 \\ \hline
\textbf{Normal/S4} & 0.57 & 0.46 & 0.053 & 0\\ \hline
\end{tabular}
\caption{Average dummy reads per data access}
\label{table:dummyratio}
\vspace{-5mm}
\end{table}

\subsection{Traffic Reduction}
The  advantage of  LAORAM is that the design guarantees that all the blocks in a superblock are going to be accessed after a single path is fetched. Comparing fat-tree and normal tree implementation of LAORAM, the bandwidth requirement of fat-tree increases compared to normal tree by a factor $\frac{3Z+1}{2(Z+1)}$, where $Z$ is the bucket size. Incorporating the increase in bandwidth because of fat-tree and the decrease in bandwidth because of superblocks, we can show that the upper bound on the bandwidth reduction for LAORAM with fat-tree compared to baseline PathORAM is $\frac{2(Z+1)}{3Z+1}*superblockSize$. The upper bound on the bandwidth reduction for LAORAM with normal tree compared to baseline PathORAM is $superblockSize$. 

\begin{figure}
\centering
  \includegraphics[width=8cm]{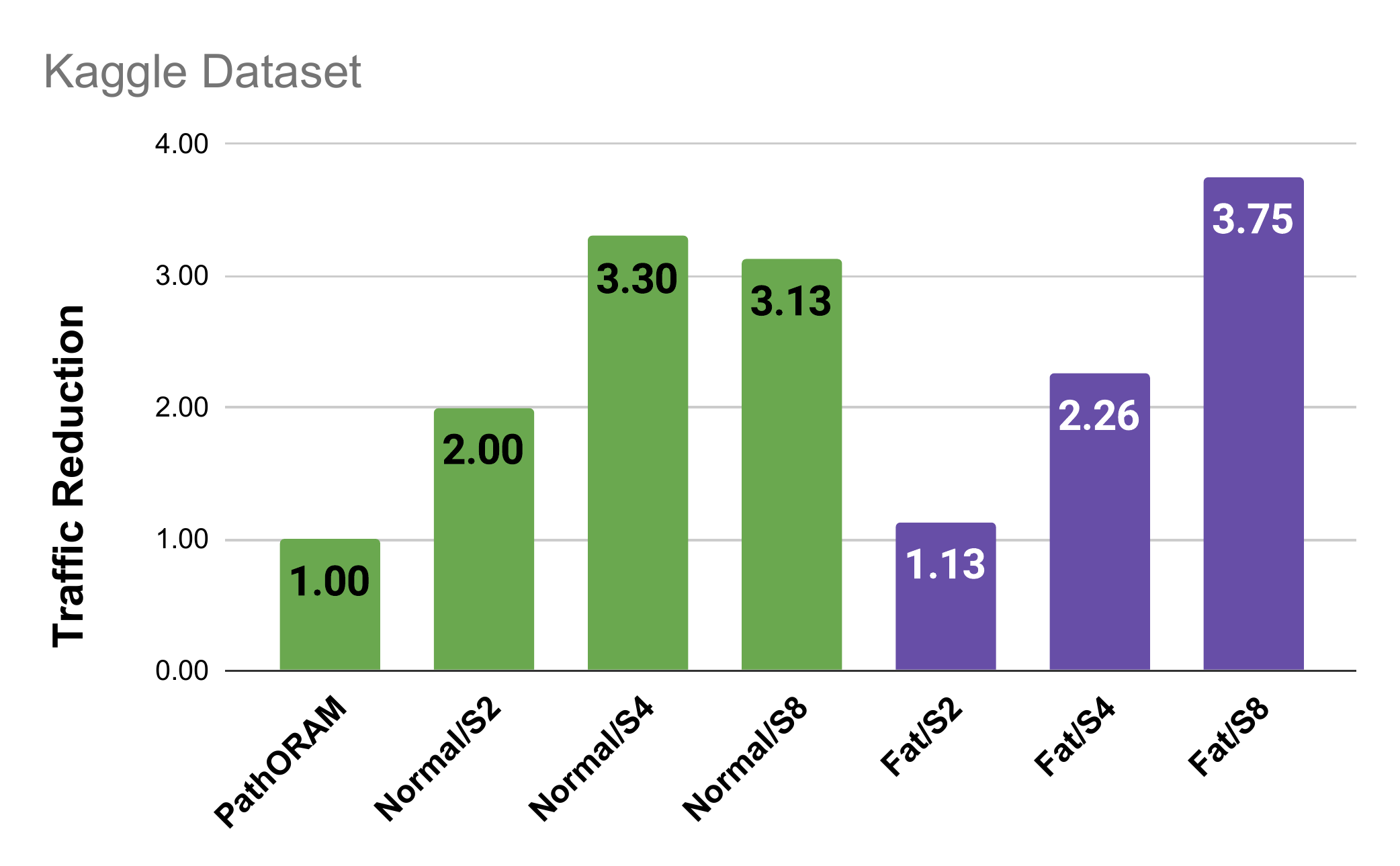}\\
  \caption{LAORAM memory traffic reduction: DLRM.}
  \label{fig:traffic-reduction}
  \vspace{-5mm}
\end{figure}

Figure \ref{fig:traffic-reduction} shows the measured bandwidth reductions with DLRM. As shown in the figure, for normal tree with superblock size of 2, the bandwidth reduction is 2 times which is equal to our theoretical bound. Theoretical bounds match the empirical data  when the number of background evictions is 0. For superblock size 4, the number of background evictions start effecting the bandwidth reduction and hence the bandwidth reduction for normal tree with superblock size 4 is 3.30X which is lower than the theoretical upper bound savings of 4X. Same observations apply to normal tree with superblock size of 8. 

The bandwidth reduction for fat-tree for smaller superblocks is less than the normal tree which is in line with our expectation as fat-tree is not useful when the number of background evictions is very low. The bandwidth reduction by using $fat/S8$ is more than that of $normal/S8$. This is due to the fact that even though theoretically a single access to fat-tree requires about 50\% more bandwidth compared to normal tree, the number of background evictions in fat-tree is very low in comparison to normal tree for large superblock size. We did the same analysis with permutation dataset which has higher stash contention. Our results show that fat-tree in fact sees a higher traffic reduction factor than normal tree. 

\subsection{Comparisons with Ring ORAM}
RingORAM \cite{ring-oram} is an alternative approach to reduce bandwidth and improve the runtime of PathORAM. RingORAM
only reads one block per node read. Theoretically, it reduces the bandwidth by the bucket size. This approach is orthogonal to LAORAM. LAORAM superblocks can be adopted to RingORAM as well. Instead of fetching $n \times log(N)$ data blocks from $n$ paths for every $n$ accesses, with LAORAM, only $[n \times log(N)] / S + S$ blocks from $n / S$ paths needs fetching, where $S$ is the superblock size. RingORAM that uses LAORAM's superblock formation strategy will face a similar stash overflow conundrum as LAORAM on PathORAM. Our fat-tree structure, a tree structure better-utilizing memory space, can help reduce the number of evictions needed. Thus, we believe LAORAM speedups, when implemented on RingORAM, are comparable with that of LAORAM on PathORAM. 

\section{Related Work}
\label{sec:related-work}
\textbf{Dynamic Superblocks:} A further enhancement was proposed in  PrORAM \cite{proram}, which dynamically creates and destroys superblocks using counter based history capture scheme. LAORAM relies on future access knowledge as opposed to history based predictions to provide substantial performance and bandwidth benefits. LAORAM can merge any data blocks into a superblock, while PrORAM can only merge consecutive data blocks. \textbf{Ring Based ORAM} \cite{ring-oram} has been discussed in the section~\ref{sec:evaluation}.  In String ORAM\cite{streamline-ring-oram} a more compact ORAM structure is proposed by leveraging real blocks to obfuscate memory rather than dummy blocks. 
String ORAM works closely in conjunction with DRAM scheduler to achieve its benefits, while our approach is agnostic to DRAM scheduler knowledge.  \textbf{Range access ORAMs:} Lite-rORAM and Hybrid-rORAM are two ORAMs proposed in \cite{multi-range-oram}. They are aimed to increase storage and access efficiencies for multi-range accesses. Our work primarily aims to improve the performance of training recommendation  models, where inputs are more randomized. \textbf{InvisiMem} \cite{invisimem} and \textbf{ObfusMem} \cite{obfusmem} use the logic layer in Micron's HMC architecture~\cite{HMC} to provide memory obfuscation and reduce the overhead of issuing multiple dummy accesses per real access. 
This approach requires substantial modification to memory hardware and interface, and are built on top of HMC architecture. LAORAM in comparison assumes that the memory device i.e. server\_storage is insecure and uses traditional DRAM design.
\textbf{Fork Path} \cite{fork-path} improves ORAM with a better request scheduler to merge path requests. Similarly, CP-ORAM \cite{cp-oram} also designs a special scheduler to accommodate different types of memory requests. Our work does not change any memory request scheduler. 

\section{Conclusion}
\label{sec:conclusion}
In this paper, we present LAORAM, a novel approach to take advantage of the  model training where accesses to embedding table memory blocks are known beforehand. LAORAM uses the \textit{future} knowledge to create large superblocks, which are then assigned to the same path in a PathORAM to improve access efficiency.  
LAORAM presents a novel architecture with three components, the server\_storage, the trainer\_GPU, and the preprocessor. The preprocessor processes the future access patterns and forms superblocks consisting of data blocks that are accessed together and assigns them to the same path. The LAORAM client then uses this preprocessor-generated metadata to manage the path assignments.  Mitigating the stash overflow while using superblocks, a fat-tree structure with variable bucket size is proposed instead of the usual binary tree with fixed bucket size. With the improvements mentioned above, LAORAM provides nearly 5X performance improvement compared to PathORAM on two different ML models and different data access patterns.


\bibliographystyle{IEEEtranS}
\bibliography{refs.bib}
\end{document}